\documentclass[10pt,a4paper]{article}
\usepackage[utf8]{inputenc}
\usepackage{booktabs} 
\usepackage{physics}
\usepackage{siunitx}
\usepackage{floatpag}
\usepackage{caption}
\floatpagestyle{empty} 
\usepackage{jcappub_modified}
\usepackage[open]{bookmark} 
\usepackage{xurl} 

\newcommand{\Msolar}{\ensuremath{M_\odot}}

\newcommand{\planck}{\mathrm{Pl}}
\newcommand{\MP}{\ensuremath{M_\planck}}
\newcommand{\mP}{\ensuremath{m_\planck}}
\newcommand{\lP}{\ensuremath{l_\planck}}
\newcommand{\tP}{\ensuremath{t_\planck}}
\newcommand{\TP}{\ensuremath{T_\planck}}
\newcommand{\kB}{\ensuremath{k_\mathrm{B}}}
\newcommand{\webpage}{\url{www.eemelitomberg.net/NaturalUnits}}

\title{Unit conversions and collected numbers in cosmology}

\date{Version 2.0}

\author{Eemeli Tomberg}

\affiliation{Laboratory of High Energy and Computational Physics, National Institute of Chemical Physics and Biophysics, R\"avala pst.~10, 10143 Tallinn, Estonia}

\emailAdd{eemeli.tomberg@kbfi.ee}

\abstract{This note collects together useful unit conversions and numerical values from early universe cosmology. It is a quick reference that can be used to make easy order-of-magnitude estimates. Included are tables for unit conversions, the thermal history of the universe, and collected properties of astronomical objects. The note also introduces a modifiable Mathematica package \texttt{NaturalUnits} (newest version \texttt{2.0}), which makes it easy to convert between natural and physical units.}

\begin{document}

\maketitle

\section{Introduction}
\label{sec:intro}


This note contains collections of numerical quantities, constants, and unit conversions useful in cosmology and particle physics. It is meant to be an easy-to-use reference for these quantities and for making order-of-magnitude estimates, with an emphasis on early universe phenomenology. Example uses include  estimating the masses of primordial black holes and the frequencies of gravitational waves originating from different eras, or estimating the number density of dark matter given the particle mass. The note comes together with an easy-to-use and modifiable Mathematica package \texttt{NaturalUnits} for unit conversions, available for download at \webpage.

The note is organized as follows: Section \ref{sec:tables} gives conversion tables for various units, including their conversions into powers of electronvolts in natural units ($c=\hbar=\kB=1$), a table of useful numbers during the history of the universe, and a table of the properties of some astronomical objects. Section \ref{sec:mathematica} introduces the associated Mathematica package.

This note and the Mathematica package are by no means comprehensive and may be updated in the future. If you have suggestions for improvements, please contact the author.

\section{Tables}
\label{sec:tables}

This section contains tables of unit conversions for energy, mass, distance, time, and others on page \pageref{tab:energy}. The unit on the left is given in terms of the unit at the top. Conversions to electronvolts (eV) are also provided, in natural units where the speed of light $c$, the reduced Planck constant $\hbar$, and the Boltzmann constant $\kB$ are set to one. Units: J = Joule, K = Kelvin, $\TP$~= Planck temperature~=~$\sqrt{\hbar c^5/(G \kB^2)}$ ($G$ = Newton's constant), eV = electronvolt, kg = kilogram, $\Msolar$ = the solar mass, $\mP$ = Planck mass = $\sqrt{\hbar c/G}$, $\MP$ = reduced Planck mass = $\mP / \sqrt{8\pi}$, m = meter, pc = parsec, AU = astronomical unit, $\lP$ = Planck length = $\sqrt{\hbar G/c^3}$, s = second, y = year, $\tP$ = Planck time = $\sqrt{\hbar G/c^5}$, Hz = Hertz.

The table on page \pageref{tab:history} presents the thermal history of the universe. Standard $\Lambda$CDM cosmology with quasi-de Sitter slow-roll inflation and instantaneous reheating is assumed, with no visible matter beyond the SM. The numerical values needed to derive these are mostly taken from the textbook by Liddle and Lyth \cite{Lyth:2009zz}; for other modern textbooks, see \cite{Mukhanov:2005sc, Weinberg:2008zzc}, and for useful lecture notes, see \cite{Baumann:2018muz, HelsinkiCosmologyNotes}. More accurate parameter values can be found e.g. in \cite{Aghanim:2018eyx, Zyla:2020zbs}. Explanations of symbols: $\rho=$~energy density, $H=$~Hubble parameter, $k_*=$~comoving scale of the Hubble radius, $T=$~temperature, $z=$~redshift, $g_*=$~the effective number of degrees of freedom, $t=$~the age of the universe, $M=$~mass within one Hubble radius, $\Omega_X=$~the energy density fraction of dark energy ($\Lambda$), cold dark matter~(CDM), baryons~(b), and radiation~(r), $\Lambda=$~cosmological constant.

The table on page \pageref{tab:astronomy} lists the properties of some astronomical objects and other observables, with $M=$~mass, $R=$~radius, $N=$~number count, $\rho=$~density, $v=$~velocity, $R_S=$~Schwarzschild radius. For more accurate values, see e.g. \cite{Zyla:2020zbs,10.1093/mnras/stw2759,Watkins_2019}.

\begin{table}[p]
\label{tab:energy}
\hypertarget{hyp:energy}{}
\bookmark[rellevel=1,keeplevel,dest=hyp:energy]{Energy conversion}
\begin{center}
\begin{tabular}{l|rrrr}
 & J & K & $\TP$ & eV \\ \midrule
J & -- & $7.24 \times 10^{22}$ & $5.11 \times 10^{-10}$ & $6.24 \times 10^{18}$ \\ 
K & $1.38 \times 10^{-23}$ & -- & $7.06 \times 10^{-33}$ & $8.62 \times 10^{-5}$ \\ 
$\TP$ & $1.96 \times 10^{9}$ & $1.42 \times 10^{32}$ & -- & $1.22 \times 10^{28}$ \\ 
eV & $1.6 \times 10^{-19}$ & $1.16 \times 10^{4}$ & $8.19 \times 10^{-29}$ & -- \\ 
\end{tabular}
\end{center}
\end{table}

\begin{table}[p]
\label{tab:mass}
\hypertarget{hyp:mass}{}
\bookmark[rellevel=1,keeplevel,dest=hyp:mass]{Mass conversion}
\begin{center}
\begin{tabular}{l|rrrrr}
 & kg & $\Msolar$ & $\mP$ & $\MP$ & eV \\ \midrule 
kg & -- & $5.03 \times 10^{-31}$ & $4.59 \times 10^{7}$ & $2.3 \times 10^{8}$ & $5.61 \times 10^{35}$ \\ 
$\Msolar$ & $1.99 \times 10^{30}$ & -- & $9.14 \times 10^{37}$ & $4.58 \times 10^{38}$ & $1.12 \times 10^{66}$ \\ 
$\mP$ & $2.18 \times 10^{-8}$ & $1.09 \times 10^{-38}$ & -- & $5.01$ & $1.22 \times 10^{28}$ \\ 
$\MP$ & $4.34 \times 10^{-9}$ & $2.18 \times 10^{-39}$ & $0.20$ & -- & $2.44 \times 10^{27}$ \\ 
eV & $1.78 \times 10^{-36}$ & $8.96 \times 10^{-67}$ & $8.19 \times 10^{-29}$ & $4.11 \times 10^{-28}$ & -- \\ 
\end{tabular}
\end{center}
\end{table}

\begin{table}[p]
\label{tab:length}
\hypertarget{hyp:length}{}
\bookmark[rellevel=1,keeplevel,dest=hyp:length]{Length conversion}
\begin{center}
\begin{tabular}{l|rrrrr}
 & m & pc & AU & $\lP$ & eV${}^{-1}$ \\ \midrule 
m & -- & $3.24 \times 10^{-17}$ & $6.68 \times 10^{-12}$ & $6.19 \times 10^{34}$ & $5.07 \times 10^{6}$ \\ 
pc & $3.09 \times 10^{16}$ & -- & $2.06 \times 10^{5}$ & $1.91 \times 10^{51}$ & $1.56 \times 10^{23}$ \\ 
AU & $1.5 \times 10^{11}$ & $4.85 \times 10^{-6}$ & -- & $9.26 \times 10^{45}$ & $7.58 \times 10^{17}$ \\ 
$\lP$ & $1.62 \times 10^{-35}$ & $5.24 \times 10^{-52}$ & $1.08 \times 10^{-46}$ & -- & $8.19 \times 10^{-29}$ \\ 
eV${}^{-1}$ & $1.97 \times 10^{-7}$ & $6.39 \times 10^{-24}$ & $1.32 \times 10^{-18}$ & $1.22 \times 10^{28}$ & -- \\ 
\end{tabular}
\end{center}
\end{table}

\begin{table}[p]
\label{tab:time}
\hypertarget{hyp:time}{}
\bookmark[rellevel=1,keeplevel,dest=hyp:time]{Time conversion}
\begin{center}
\begin{tabular}{l|rrrr}
 & s & y & $\tP$ & eV${}^{-1}$ \\ \midrule 
s & -- & $3.17 \times 10^{-8}$ & $1.85 \times 10^{43}$ & $1.52 \times 10^{15}$ \\ 
y & $3.16 \times 10^{7}$ & -- & $5.85 \times 10^{50}$ & $4.79 \times 10^{22}$ \\ 
$\tP$ & $5.39 \times 10^{-44}$ & $1.71 \times 10^{-51}$ & -- & $8.19 \times 10^{-29}$ \\ 
eV${}^{-1}$ & $6.58 \times 10^{-16}$ & $2.09 \times 10^{-23}$ & $1.22 \times 10^{28}$ & -- \\ 
\end{tabular}
\end{center}
\end{table}

\begin{table}[p]
\label{tab:other}
\hypertarget{hyp:other}{}
\bookmark[rellevel=1,keeplevel,dest=hyp:other]{Other conversion}
\begin{center}
\begin{tabular}{l|rrrr}
 & Hz & Mpc${}^{-1}$ & $\MP$ & GeV \\ \midrule 
Hz & -- & $1.03 \times 10^{14}$ & $2.7 \times 10^{-43}$ & $6.58 \times 10^{-25}$ \\ 
Mpc${}^{-1}$ & $9.72 \times 10^{-15}$ & -- & $2.63 \times 10^{-57}$ & $6.39 \times 10^{-39}$ \\ 
$\MP$ & $3.7 \times 10^{42}$ & $3.81 \times 10^{56}$ & -- & $2.44 \times 10^{18}$ \\ 
GeV & $1.52 \times 10^{24}$ & $1.56 \times 10^{38}$ & $4.11 \times 10^{-19}$ & -- \\ 
\end{tabular}
\end{center}
\end{table}

\begin{table}[p]
\label{tab:history}
\hypertarget{hyp:history}{}
\bookmark[rellevel=1,keeplevel,dest=hyp:history]{Thermal history}
\begin{center}
\begin{tabular}{l} \toprule
Cosmic inflation (r=tensor-to-sclar ratio) \\
$\rho^{1/4} = \qty(\frac{r}{0.08})^{1/4}\times \SI{1.7e16}{GeV}$ \\
$H = \qty(\frac{r}{0.08})^{1/2}\times \SI{7e13}{GeV} = \qty(\frac{r}{0.08})^{1/2} \times \SI{e52}{Mpc^{-1}}$ \\
$k_* = \SI{0.05}{Mpc^{-1}} = \SI{5e-16}{Hz}$ (CMB scale) \\
\\
\midrule
Reheating \\
$z= 10^{29 + \frac{1}{4}\log_{10} \frac{r}{0.08}} = e^{67 + \frac{1}{4}\ln \frac{r}{0.08}}$ \\
$k_* = 10^{23 + \frac{1}{4}\log_{10} \frac{r}{0.08}} \si{Mpc^{-1}} = 10^{9 + \frac{1}{4}\log_{10} \frac{r}{0.08}} \si{Hz}$ \\ \\
\midrule
Radiation domination: $a \propto t^{1/2}$, $H = 1/(2t) \propto a^{-2}$ \\
\midrule
Electroweak phase transition \\
\begin{tabular} {@{}lll}
$T = \SI{100}{GeV}$ & $z= 10^{15}=e^{35}$ & $H = \SI{e-5}{eV} = \SI{e24}{Mpc^{-1}}$ \\
$g_* = 106.75$ & $t= \SI{20}{ps}$ & $M= \num{e-6} \Msolar = \SI{e28}{g}$ \\
$k_* = \SI{e9}{Mpc^{-1}} = \SI{e-5}{Hz}$
\end{tabular} \\ \\
\midrule
QCD phase transition \\
\begin{tabular}{@{}lll}
$T = \SI{150}{MeV}$ & $z= 10^{12}=e^{28}$ & $H = \SI{e-11}{eV} = \SI{e18}{Mpc^{-1}}$ \\
$g_* = 17.25$ & $t= \SI{30}{\micro s} $ & $M= 5 \Msolar = \SI{e34}{g}$ \\
$k_* = \SI{e6}{Mpc^{-1}} = \SI{e-8}{Hz}$ \\
\end{tabular} \\ \\
\midrule
Big Bang Nucleosynthesis \\
\begin{tabular}{@{}lll}
$T = \SI{100}{keV}$ & $z= 10^{8}=e^{20}$ & $H = \SI{e-18}{eV} = \SI{e11}{Mpc^{-1}}$ \\
$g_* \approx 3$ & $t= \SI{100}{s} = \SI{2}{min}$ & $M= \num{e7} \Msolar = \SI{e41}{g}$ \\
$k_* = \SI{e3}{Mpc^{-1}} = \SI{e-11}{Hz}$ \\
\end{tabular} \\ \\
\midrule
Matter-radiation equality \\
\begin{tabular}{@{}lll}
$T = \SI{0.8}{eV} = \SI{9000}{K}$ & $z= \num{3e3}=e^{8}$ & $H = \SI{30}{Mpc^{-1}}$ \\
$t= \SI{60 000}{y} = \SI{2e12}{s} $ & $M= \num{e17} \Msolar = \SI{e51}{g}$ \\
$k_* = \SI{e-2}{Mpc^{-1}} = \SI{e-16}{Hz}$ \\
\end{tabular} \\ \\
\midrule
Matter domination: $a \propto t^{2/3}$, $H = 2/(3t) \propto a^{-3/2}$ \\
\midrule
Recombination \\
\begin{tabular}{@{}lll}
$T = \SI{0.25}{eV} = \SI{3000}{K}$ & $z= 1050=e^{7}$ & $H = \SI{6}{Mpc^{-1}}$ \\
$t = \SI{400000}{y} = \SI{e13}{s} $ & $M= \num{e18} \Msolar$ \\
$k_* = \SI{5e-3}{Mpc^{-1}} = \SI{5e-17}{Hz}$ \\
\end{tabular} \\ \\
\midrule
Today \\
\begin{tabular}{@{}lll}
$T = \SI{2.7}{K} = \SI{0.23}{meV}$ & $t= \SI{13.8e9}{y} = \SI{4.35e17}{s}$ & $M= 10^{22} \Msolar$
\end{tabular} \\
\begin{tabular}{@{}llll}
$\Omega_\Lambda = 0.72$ & $\Omega_\text{CDM} = 0.23$ & $\Omega_\text{b} = 0.05$ & $\Omega_\text{r} = \num{8e-5}$
\end{tabular} \\
$H = \SI{70}{km/s/Mpc} = \SI{2e-4}{Mpc^{-1}} = 1/(\SI{4000}{Mpc}) = \SI{e-33}{eV} = \SI{e-61}{\MP}$ \\
$\rho = \SI{9e-27}{kg/m^3} = \SI{5}{GeV/m^3} = 10^{-120} \MP^4$ \\
$\Lambda = \SI{e-120}{\MP^2} = \SI{e-52}{m^{-2}} = 1/(\SI{3000}{Mpc})^2$ \\ \\
\bottomrule
\end{tabular}
\end{center}
\end{table}

\begin{table}
\label{tab:astronomy}
\hypertarget{hyp:astronomy}{}
\bookmark[rellevel=1,keeplevel,dest=hyp:astronomy]{Astronomical quantities}
\begin{center}
\begin{tabular}{l}
\toprule
Milky Way \\
$M \sim \SI{e12}{\Msolar}$ (mostly dark matter) \\
$M_\text{stars} \sim \SI{5e10}{\Msolar}$ \quad
$R \sim \SI{e2}{kpc}$ \quad $N_\mathrm{stars} \sim 10^{11}$ \\
\midrule
Local dark matter halo \\
$\rho \sim \SI{0.3}{GeV/cm^3}$ \qquad
$v \sim \SI{200}{km/s}$ \\
\midrule
Sun \\
$M = \SI{2e30}{kg}$ \qquad
$R = \SI{7e8}{m}$ \qquad
$R_S = \SI{3}{km}$ \\
\midrule
Earth \\
$M = \SI{6e24}{kg}$ \qquad
$R = \SI{6400}{km}$ \qquad
$R_S = \SI{9}{mm}$  \\
\midrule
Moon \\
$M = \SI{7e22}{kg}$ \qquad
$R = \SI{1700}{km}$ \qquad
$R_S = \SI{0.1}{mm}$  \\
\bottomrule
\end{tabular}
\end{center}
\end{table}

\clearpage
\section{Mathematica package \texttt{NaturalUnits}, version \texttt{2.0}}
\label{sec:mathematica}
The Mathematica package for unit conversions, \texttt{NaturalUnits}, can be downloaded from the author's web page, \webpage. This section briefly describes the functionality of the package's version \texttt{2.0}; more examples and tutorials can be found on the web page.

To use the package, download \texttt{NaturalUnits.m} and place it into the same folder with your Mathematica notebook. Load the package and set up the natural units in the notebook with the commands
\begin{verbatim}
In:  SetDirectory[NotebookDirectory[]];
     << "NaturalUnits.m"
     NaturalUnitsSetup[NewUnitSystem->CosmologyUnits, NewNatUnits->{eV}];
Out: NaturalUnits 2.0
\end{verbatim}
The last line sets up the natural unit system used in this note, $c=\hbar=\kB=1$, with electronvolts as the default base unit. In version \texttt{2.0}, other unit systems are also supported, and custom systems can be set up by giving the defining equations, e.g. \texttt{NewUnitSystem->\{c==G==1\}} in the above for geometrized units.

To do unit conversions, call the function \texttt{NaturalConvert}. For example,
\begin{verbatim}
In:  NaturalConvert[10^-4 MSolar, kg]
Out: 1.989*10^26 kg
\end{verbatim}
The first parameter is the quantity to be converted, with units represented by Mathematica symbols. The second parameter is the target unit. If the second parameter is left empty, a conversion to a power of the base units is made:
\begin{verbatim}
In:  NaturalConvert[10^-4 MSolar]
Out: 1.11575*10^62 eV
\end{verbatim}
If the two units don't match in their natural-unit form (powers of electronvolts), a conversion to the correct power is attempted for the target unit. Each input to \texttt{NaturalConvert} must be proportional to a single power of the natural units, otherwise the conversion fails.

The package supports many units useful for cosmology, together with prefixes such as GeV~$=10^9$~eV, and also knows some common constants of nature, corresponding to standard symbols when possible\footnote{For less obvious special cases, \texttt{MSolar}, \texttt{MEarth}, and \texttt{MMoon} are the masses of the Sun, the Earth and the Moon, \texttt{mElectron} and \texttt{mProton} are the electron and proton masses, and the Planck units are given by \texttt{mP} (Planck mass), \texttt{MP} (reduced Planck mass), \texttt{tP} (Planck time), \texttt{lP} (Planck length), \texttt{TP} (Planck temperature), and \texttt{EP} (Planck energy).}. A full list of symbols is contained in \texttt{\$AllUnits}, and can be printed together with unit definitions with \texttt{NaturalUnitsList[]}. Adding custom units is simple: open \texttt{NaturalUnits.m} and add the unit to the list \texttt{\$UnitRules} in terms of the previously defined units and constants, like the other entries in the list. New units can also be added during a notebook session with the command \texttt{NaturalUnitsAdd}, e.g.
\begin{verbatim}
In:  NaturalUnitsAdd[mile->1.609344 km]
     NaturalConvert[10 mile/h, m/s]
Out: 4.4704 m/s
\end{verbatim}

For more advanced functionality, the second parameter to \texttt{NaturalConvert} can be a list of units $\{x,y,z,\dots\}$, and the function then tries to convert the quantity into a product of powers of these units, $x^{n_1}\times y^{n_2}\times z^{n_3}\dots$ . Option \texttt{KeepUnit} tells whether the units should be kept in the result (\texttt{KeepUnit->True}, default), kept in a non-simplified form (\texttt{KeepUnit->Defer}), or dropped (\texttt{KeepUnit->False}), for example:
\begin{verbatim}
In:  NaturalConvert[200 ly/rad, {pc, 180 deg}, KeepUnit->Defer]
Out: 192.639 pc/(180 deg)
\end{verbatim}
The default value of the target unit is stored in the option \texttt{ToUnitDefault} (its starting value is \texttt{NoUnit}, which produces the above-described results in powers of eV).

Beware: since many simple symbols are used as units, there may be overlap with other packages and custom code. The package is best used on its own for simple number manipulation; otherwise, the namespace label \texttt{NaturalUnits`} should be used in front of the unit symbols.

\paragraph{Examples of use.} The package can be used to easily calculate quantities that would otherwise require cumbersome unit conversions. For example, the Schwarzschild radius of a 10 solar mass black hole:
\begin{verbatim}
In:  NaturalConvert[2*10 MSolar*G, km]
Out: 29.5399 km
\end{verbatim}
The typical frequency of CMB radiation, starting from temperature:
\begin{verbatim}
In:  NaturalConvert[2.7 K, Hz]
Out: 3.53485*10^11 Hz
\end{verbatim}
Typical mass of a primordial black hole formed when the temperature of the universe was \SI{100}{MeV} (that is, mass of a sphere with radius $H^{-1}$, with $\rho \sim T^4$, $H^2 \MP^2 \sim \rho$):
\begin{verbatim}
In:  NaturalConvert[(T^4*(T^2/MP)^-3) /. T -> 100 MeV, MSolar]
Out: 1.29459 MSolar
\end{verbatim}
More examples can be found at \webpage.


\acknowledgments
This work was supported by the Estonian Research Council grants PRG803, PRG1055, and MOBTT5 and by the EU through the European Regional Development Fund CoE program TK133 ``The Dark Side of the Universe''.

\bibliographystyle{JHEP}
\bibliography{conv}

\providecommand{\href}[2]{#2}\begingroup\raggedright\begin{thebibliography}{1}

\bibitem{Lyth:2009zz}
D.~H. Lyth and A.~R. Liddle, \emph{{The primordial density perturbation:
  Cosmology, inflation and the origin of structure}}.
\newblock 2009.

\bibitem{Mukhanov:2005sc}
V.~Mukhanov, \emph{{Physical Foundations of Cosmology}}.
\newblock Cambridge University Press, Oxford, 2005.

\bibitem{Weinberg:2008zzc}
S.~Weinberg, \emph{{Cosmology}}.
\newblock 2008.

\bibitem{Baumann:2018muz}
D.~Baumann, \emph{{Primordial Cosmology}},
  \href{http://dx.doi.org/10.22323/1.305.0009}{\emph{PoS} {\bfseries TASI2017}
  (2018) 009}, [\href{https://arxiv.org/abs/1807.03098}{{\ttfamily
  1807.03098}}].

\bibitem{HelsinkiCosmologyNotes}
H.~Kurki-Suonio, ``{Cosmology I and II}.'' Available at
  \url{https://www.mv.helsinki.fi/home/hkurkisu/}, referenced in January 2021.

\bibitem{Aghanim:2018eyx}
{\scshape Planck} collaboration, N.~Aghanim et~al., \emph{{Planck 2018 results.
  VI. Cosmological parameters}},
  \href{http://dx.doi.org/10.1051/0004-6361/201833910}{\emph{Astron.
  Astrophys.} {\bfseries 641} (2020) A6},
  [\href{https://arxiv.org/abs/1807.06209}{{\ttfamily 1807.06209}}].

\bibitem{Zyla:2020zbs}
{\scshape Particle Data Group} collaboration, P.~Zyla et~al., \emph{{Review of
  Particle Physics}}, \href{http://dx.doi.org/10.1093/ptep/ptaa104}{\emph{PTEP}
  {\bfseries 2020} (2020) 083C01}.

\bibitem{10.1093/mnras/stw2759}
P.~J. McMillan, \emph{{The mass distribution and gravitational potential of the
  Milky Way}}, \href{http://dx.doi.org/10.1093/mnras/stw2759}{\emph{Monthly
  Notices of the Royal Astronomical Society} {\bfseries 465} (10, 2016)
  76--94},
  [\href{https://arxiv.org/abs/https://academic.oup.com/mnras/article-pdf/465/1/76/8593676/stw2759.pdf}{{\ttfamily
  https://academic.oup.com/mnras/article-pdf/465/1/76/8593676/stw2759.pdf}}].

\bibitem{Watkins_2019}
L.~L. Watkins, R.~P. van~der Marel, S.~T. Sohn and N.~W. Evans, \emph{Evidence
  for an intermediate-mass milky way from {Gaia DR2} halo globular cluster
  motions}, \href{http://dx.doi.org/10.3847/1538-4357/ab089f}{\emph{The
  Astrophysical Journal} {\bfseries 873} (mar, 2019) 118}.

\end{thebibliography}\endgroup

\end{document}